\documentclass[sigconf]{acmart}
\AtBeginDocument{%
  \providecommand\BibTeX{{%
    \normalfont B\kern-0.5em{\scshape i\kern-0.25em b}\kern-0.8em\TeX}}}

\setcopyright{rightsretained}
\copyrightyear{2023}
\acmYear{2023}
\acmDOI{XXXXXXX.XXXXXXX}

\acmConference[]{Feb 9}{2023}{Pre-print submitted to arXiv}
\acmBooktitle{Woodstock '18: ACM Symposium on Neural Gaze Detection,
 June 03--05, 2018, Woodstock, NY} 
\acmPrice{15.00}
\acmISBN{978-1-4503-XXXX-X/18/06}

\usepackage{float}
\begin{document}

\title{Guttation Monitor: Wearable Guttation Sensor for Plant Condition Monitoring and Diagnosis}


\author{Qiuyu Lu}
\orcid{0000-0002-8499-3091}
\affiliation{%
  \institution{Carnegie Mellon University}
  \streetaddress{5000 Forbes Ave}
  \city{Pittsburgh}
  \state{PA}
  \country{USA}
  \postcode{15289}
}
\email{qiuyul@cs.cmu.edu}

\author{Lydia Yang}
\affiliation{%
  \institution{Carnegie Mellon University}
  \streetaddress{5000 Forbes Ave}
  \city{Pittsburgh}
  \state{PA}
  \country{USA}}
\email{llyang@andrew.cmu.edu}

\author{Aditi Maheshwari}
\affiliation{%
  \institution{Accenture Labs}
  \streetaddress{}
  \city{San Francisco}
  \state{CA}
  \country{USA}}
\email{aditi.maheshwari@accenture.com}

\author{Hengrong Ni}
\affiliation{%
  \institution{Carnegie Mellon University}
  \city{Pittsburgh}
  \state{PA}
  \country{USA}}
\email{hni01@alumni.risd.edu}

\author{Tianyu Yu}
\affiliation{%
  \institution{Tsinghua University}
  \streetaddress{}
  \city{Beijing}
  \state{}
  \country{China}}
 \email{tyy21@mails.tsinghua.edu.cn}

 \author{Jianzhe Gu}
\affiliation{%
  \institution{Carnegie Mellon University}
  \streetaddress{5000 Forbes Ave}
  \city{Pittsburgh}
  \state{PA}
  \country{USA}}
\email{jianzheg@andrew.cmu.edu}

\author{Advait Wadhwani}
\affiliation{%
  \institution{Carnegie Mellon University}
  \streetaddress{5000 Forbes Ave}
  \city{Pittsburgh}
  \state{PA}
  \country{USA}}
\email{advaitw@andrew.cmu.edu}

\author{Andreea Danielescu}
\affiliation{%
  \institution{Accenture Labs}
  \streetaddress{}
  \city{San Francisco}
  \state{CA}
  \country{USA}}
\email{andreea.danielescu@accenture.com}

\author{Lining Yao}
\affiliation{%
  \institution{Carnegie Mellon University}
  \streetaddress{5000 Forbes Ave}
  \city{Pittsburgh}
  \state{PA}
  \country{USA}}
\email{liningy@cs.cmu.edu}

\renewcommand{\shortauthors}{Lu, et al.}

\begin{abstract}
  Plant life plays a critical role in the ecosystem. However, it is difficult for humans to perceive plants’ reactions because the biopotential and biochemical responses are invisible to humans. Guttation droplets contain various chemicals which can reflect plant physiology and environmental conditions in real time. Traditionally, these droplets are collected manually and analyzed in the lab with expensive instruments. Here, we introduce the Guttation Monitor, an on-site and low-cost monitoring technology for guttation droplets. It consists of three parts 1) a paper-based microfluidic chip that can collect guttation droplets and perform colorimetric detection of six chemicals, 2) a self-contained and solar-powered camera module that can capture the result from the chip, and 3) an end-user app that can interpret the result. \textcolor{black}{We discuss this technology's design and implementation, conduct evaluations on tomato plants, conduct interviews, and envision how such a technology could enhance the human-plant relationship in four dimensions.}
\end{abstract}



\begin{teaserfigure}
  \includegraphics[width=\textwidth]{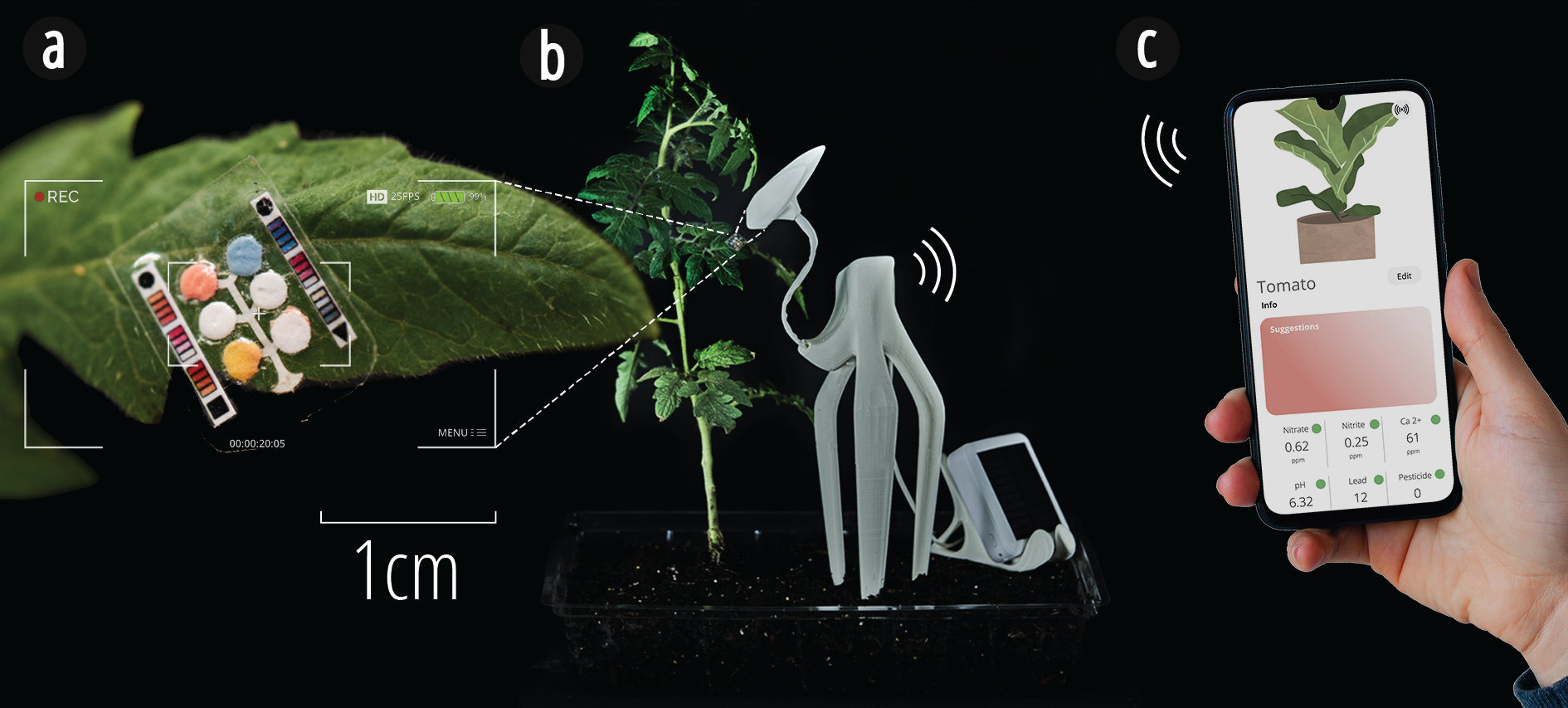}
  \caption{(a) Close view of the guttation chip on the plant. (b) The camera module captures the colorimetric detection result and sends it to the server. (c) The server performs image processes and sends the monitoring result and interpretation to the end-user application.}
  \Description{Wearable Guttation Sensor Design}
  \label{teaser}
\end{teaserfigure}

\maketitle

\section{INTRODUCTION}
Plants are a vital part of the ecosystem. They provide food sources for many organisms and play an important role in maintaining a global climate suitable for life. Furthermore, agriculture, horticulture, landscaping, and other plant-involved activities are also important to human civilization. In Human-Computer Interaction (HCI), we have seen much work focusing on creating smart, interactive systems involving plants \cite{10.1145/1240866.1241037, 10.1145/3290607.3311778}.

Wearable plant sensors are gaining popularity due to the rising demand for monitoring plant statuses in smart gardens and farms, battling climate change, and coordinating plants and humans \cite{Lee2021}. Many wearable sensors that detect environmental stressors \cite{Nassar2018, 8486710}, plant growth \cite{Tang2019, Jiang2020}, and plant volatile organic compounds (VOC) \cite{Li2021, Calisgan2020} have been developed. However, among all plant physiological activities, guttation is unique and has not been well explored yet. Guttation droplets is a common secretion of plants. It consists of various organic and inorganic chemicals that can be leveraged to understand the plant's status and environmental conditions. While traditionally, such guttation droplets are collected manually and analyzed in the lab with expensive instruments, we introduce the Guttation Monitor, an on-site and low-cost monitoring technology for guttation droplets (Fig. \ref{teaser}). The Guttation Monitor consists of 1) paper-based microfluidic chips that can collect the guttation droplet and perform colorimetric detection of chemicals (determining the concentration of a chemical with the aid of a color reagent); 2) a self-contained and solar-powered camera module that can capture the detection result of the chip; 3) and an end-user app that can perform image processing on the camera input and interpret the result. 

We briefly summarize our core contributions below. From the technical and design perspectives:
\begin{itemize}
    \item This is the first engineering system that senses guttation droplets via a plant wearable device. We have shown that six types of chemicals can be sensed on a single paper-based microfluidic chip. Though the chip is single-use, it is very low-cost \footnote{\textcolor{black}{Calcumalated based on the US Amazon rental price of the material used, the cost of the chip is around 50 cents. We also sent inquiries to suppliers and factories in Asia. Based on their quotations, the cost can be reduced to less than 10 cents.}}  
    \item The system design is uniquely tailored to plant guttation sensing. The chip is conformable, ultra-lightweight ($\sim$0.03 g), and requires a very small sample. Moreover, the whole system can work unattended for days, waiting for plants to guttate and process the result automatically. 
    \item On-plant experiments are conducted to validate the system and primarily explore how we can leverage the sensor data to interpret plants’ status and provide corresponding suggestions to humans.
\end{itemize}

In terms of the potential implication or benefits to the field of HCI, we hope:
\begin{itemize}
    \item The wearable plant sensor will become an enabling technology for researchers interested in human-plant interaction, in various contexts such as plant status monitoring, soil condition monitoring, and augmented interactions among plants, humans, and other species (e.g., pollinators), etc.
    \item The sensor, given a chance to be deployed and utilized by gardeners and farmers in the future, may help collect data across time and space, contributing to constructing digital and smart gardens.
    \item The technical components introduced in this paper, such as low-cost wearable fluidic sensors and simple yet effective imaging processing of the fluidic chips, may be leveraged by HCI researchers interested in other application domains, such as on-skin wearable sensors for humans and animals.
\end{itemize}

\section{BACKGROUND KNOWLEDGE}
\begin{figure*}[t]
    \includegraphics[width=\textwidth]{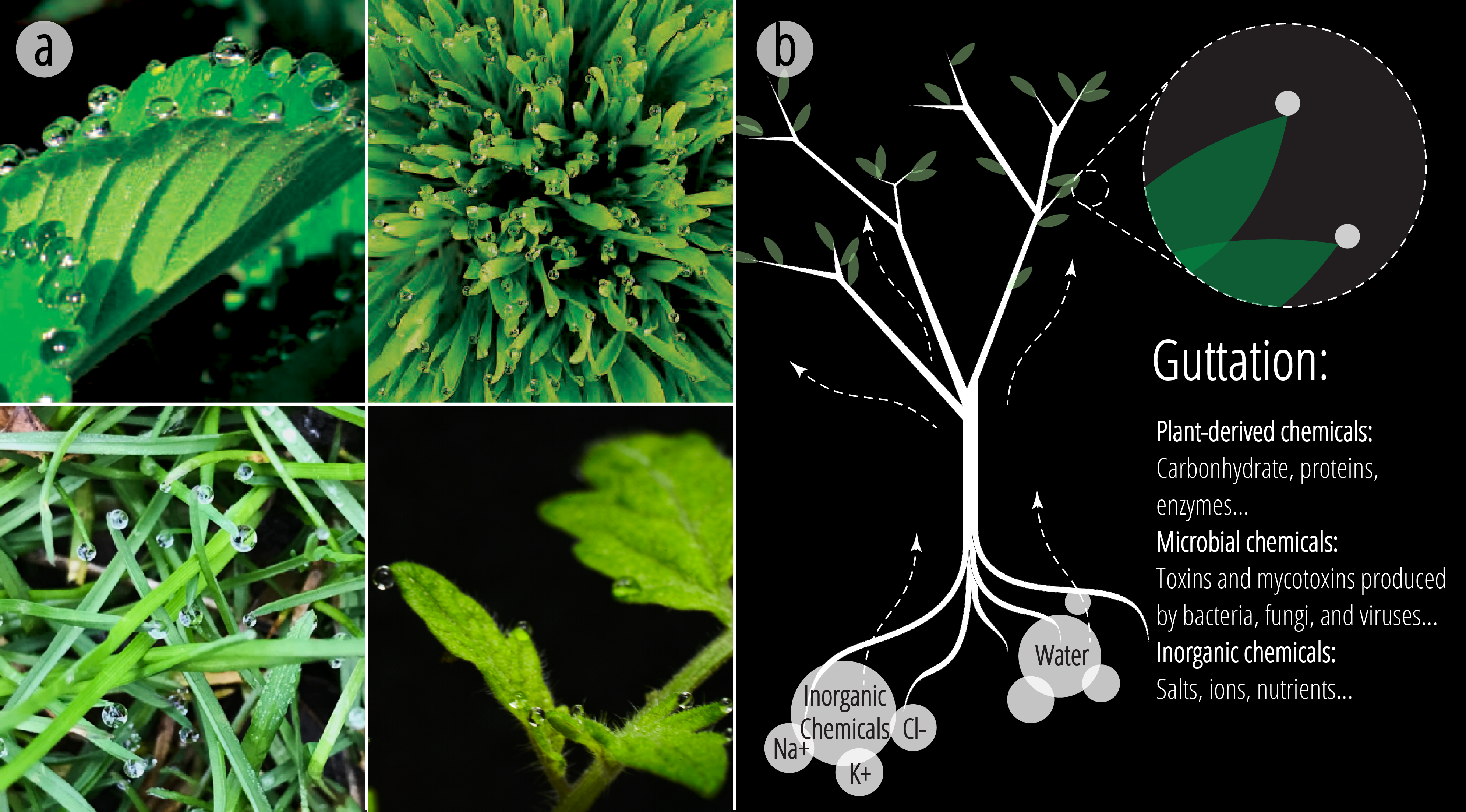}
    \caption{(a) Guttation in the form of droplets in different plant species: strawberry (source: \href{ https://commons.wikimedia.org/wiki/File:Guttation\_ne.jpg}{Wikemedia}), annual blugrass (source: \href{https://commons.wikimedia.org/wiki/File:Leaves\_with\_drops.jpg}{Wikemedia}), bentgrass (shot by author), tomato  (shot by author). (b) Chemicals that can be found in plants’ guttation droplets. }
    \label{fig:2}
\end{figure*}
At the break of dawn, known as the teardrops of plants, beads of liquid dot the edges of various flora. Secretions are common phenomena in all life forms as part of many metabolic processes. Guttation is the process of secretion of droplets expelled through the leaves of plants (Fig. \ref{fig:2}.a). It is observed in many types of plant life, including angiosperms, gymnosperms, ferns, algae, and fungi \cite{Singh2020-dt}. This clear secretion is formed as the xylem and phloem sap is excreted through pores \cite{Singh2016}. Guttating helps the plant get rid of any excess water and materials found in the plant and is formed when there is excess moisture in the soil. Root pressure causes the liquid in the veins to be pushed out through the pores, which is then squeezed out onto the tips of the leaves. Guttation should not be confused with transpiration or dew. Transpiration is released as a vapor produced as a byproduct of photosynthesis as water vapor is released through the stomata \cite{10.1104/pp.104.900213}. Guttation is generally observed when transpiration is low to make up for the water buildup. Transpiration decreases, and guttation increases with high humidity because the vapor concentration is high outside the plant. Dew occurs when atmospheric water vapor condenses onto the leaf's surface \cite{Beysens1995}. This generally occurs at night when the temperature is low and atmospheric humidity is high. Dew is composed of mostly pure water with dissolved atmospheric gasses. Dew composition is dependent on the atmosphere, while guttation is dependent on the plant.

Guttation is an important indicator of a plant's status. We chose guttation because it contains (Fig. \ref{fig:2}.b): 1) many inorganic compounds such as salts, ions, and nutrients \cite{Curtis1943, Curtis1944, Ivanoff1963, Pilanali2005}, which directly reflect what the plant absorbs. The soil conditions directly influence the inorganic materials found in guttation droplets. For example, guttation droplets can accurately indicate inorganic molecules such as nitrates \cite{Singh2013}. 2) organic materials produced by the plant such as carbohydrates \cite{Singh2020-dt, Slewinski2009}, proteins \cite{Magwa1993, Hehle2011}, and hormones \cite{Stokes1954}. 3) \textcolor{black}{microbial chemicals, such as toxins and mycotoxins produced by bacteria, fungi, and viruses \cite{Young1995, Scott2004, Gareis2007}. Such microbial chemicals can be found in the early stage of infection. Leveraging this we may treat the plant prophylactically before severe symptoms appear and plants stop guttating.}

This combination of inorganic and organic materials provides a snapshot of the chemical environment inside the plant. Besides, it allows a non-invasive procedure to analyze soil fertility and productivity, without all the soil collecting, dissolving, and filtering procedures. By analyzing the chemicals found in guttation, we can act accordingly to what the plant is deficient in.

It is also important to study guttation as an indicator of not just the plant's health but also the health of the environment around the plant. Guttation is crucial to the plant’s immune system by flushing out diseased pathogens \cite{Singh2020-dt}. There has been research on how the analysis of guttation can be used to engineer crops with higher benefits to herbivores and deter pathogens. Guttation containing certain carbohydrates and proteins can also serve as a nutrient-rich food for insects \cite{UrbanejaBernat2020}. In contrast, guttation can also be carriers of chemicals harmful to insects, such as insecticides. Research has found that neonicotinoid insecticides found in guttation can be harmful to insects such as bees \cite{Tapparo2011}. Thus, knowing and understanding the composition of guttation is crucial to the well-being of an ecosystem.

\section{RELATED WORK}
\subsection{Wearable Sensors on Plants}
\textcolor{black}{While technologies like aerial multi-spectral analysis\cite{sankaran2015field, geipel2016programmable} have been developed to monitor the plants on the macro scale and in large quantity, such technologies usually lack the capability of understanding diverse biomarkers and the micro-environment.} 

In the meanwhile, there is a lot of research focused on advancing \textcolor{black}{small sensors attached to living plants, like sensors attached to humans, they are called wearable sensors as well}. Such sensors are designed to detect environmental factors like temperature, humidity, drought for gardening and farming \cite{Nassar2018, 8486710, Wang2021}. Other wearable electronic sensors are designed for monitoring a plant’s growth \cite{Tang2019, Jiang2020}, water status \cite{Oren2017}, or movement \cite{Marx1975} within the plant. \textcolor{black}{However, these sensors are limited in their ability to provide insight into the plant's various physiological states or complex environmental conditions like the chemical composition of the soil. Instead, we aim to design a solution that can provide this type of insight}.

Recently, wearable plant \textcolor{black}{VOC sensors \cite{Li2021, Calisgan2020, Lew2020} have been developed to further monitor plant health and damage. However, they can only detect volatile signals, such as phytohormones and aromatics.} Similarly, many wearable sensors for humans have been developed that we can draw inspiration from. Among them, sensors that detect human bodily fluids such as sweat \cite{Koh2016, Zhang2019, Kwon2021} and tears \cite{Yao2011, Sempionatto2019} inspired us to think about designing such a sensor for plants. As discussed above, guttation droplets contain various chemicals that can help humans comprehensively, precisely, and deeply understand plant physiology and environmental conditions. 

Many aforementioned plant wearable sensors \textcolor{black}{are fabricated with expensive material like graphene \cite{Marx1975}, gold/silver nano-particles/wires \cite{Tang2019, Li2021, Lew2020}, and manufactured through sophisticated processes like metal sputtering \cite{Tang2019}, spin-coating \cite{Wang2021, Donald2004, Calisgan2020, Kim2019}, soft-/photo-lithography \cite{Marx1975, Lew2020, Kim2019}, oxygen plasma \cite{Lew2020}, and vapor coating \cite{Li2021, 8486710, Zhao2019}. Instead, we focus on low-cost, accessible materials and fabrication processes that make the technology more accessible to a broader range of people.}

\subsection{Guttation Analysis}
Guttation droplets can be collected in multiple ways, ranging from using blotting paper to using test tubes. Here we highlight a few of the main techniques used in academic research. Pedersen’s (1993) technique uses glass capillaries to collect guttation from aquatic plants \cite{Pedersen1993}. Komarnytsky et al. used negative pressure with a handheld pipette to individually collect guttation droplets \cite{Komarnytsky2000}. Wagner used solvent-filled micro-capillaries to collect guttation while only making contact around glands \cite{WAGNER2004}. Singh used noninvasive eco-friendly blotting paper to collect guttation using capillary action \cite{Singh2008}. Various protocols and instruments have been used to detect chemical concentrations in guttation droplets, including electric probes, spectrophotometers, and chromatography, which are designed to use photons to detect concentrations of chemicals \cite{Lederer1957-zq, Knemeyer2000, Klotz2018}.

\textcolor{black}{Above mentioned methods use specialized equipment and precise lab conditions for sensitive calculations of chemical concentrations. In contrast, the colorimetric detection we used in the Guttation Monitor is a simple, quick and more accessible way to analyze concentrations by exploiting the properties of various chemicals and their interactions with other chemicals. These colorimetric indicators function by reacting to the desired chemical, resulting in a visible color change. While plants guttate slowly and irregularly at night or in the early morning, and most people do not have access to professional chemical analysis equipment, an on-site, affordable, unattended detection could fill the gap for guttation monitoring.}

\subsection{Human-Plant Interaction}
Human-Plant Interaction (HPI) is an emerging subfield under HCI that tends to focus on living plants as materials and collaborators in the design of interaction systems \cite{Chang2022, Kuznetsov2011}. As such, there is a large body of work leveraging plants as novel interfaces facilitating communication between humans, plants, and the broader ecosystem. One body of HPI works focus on utilizing plants as sensing and display systems \cite{10.1145/1240866.1241037, Poupyrev2012, Manzella2013, 10.1145/3290607.3311778, Gentile2018, Kuribayashi2006}, either augmented with peripheral sensors \cite{10.5555/1780402.1780414, Angelini2016} or injected with conductive organic polymers and other biocompatible materials \cite{Stavrinidou2015, 10.1145/3290607.3311778, Kwak2017}, with the broader goals of using plants as input interfaces or output displays. Another set of HPI projects utilizes plants as emotion-evoking or persuasive entities \cite{Holstius2004, Degraen2021, Salem2008, antifakos2003laughinglily} and natural human companions for educational and wellness purposes \cite{10.5555/1780402.1780414, Seo2015, Botros2016, Hwang2010, Angelini2016}. Finally, with the emerging focus on multispecies HCI, there is a growing body of research centered around developing symbiotic human-plant relationships and nurturing empathy for non-human entities like plants and the environment \cite{Kobayashi2015, 10.1145/3290607.3311778, Cheok2008, 10.1145/1240866.1241037}. We build on this body of work by developing novel plant-interfacing sensor chips that can allow us to gain greater insights into plant well being by looking beyond well explored parameters like moisture content and into more versatile signals in plant fluids. Our guttation chips are biocompatible, and accessible, holding the potential to strengthen human-plant relationships by empowering plant caretakers to better understand the needs of their plants through more detailed and granular inquiry into plant health. 

\section{PLANT EPIDERMAL MICROFLUIDIC SENSOR ENABLED GUTTATION MONITORING}
Our plant-to-user Guttation Monitor System consists of 1) The Sensor, a paper-based microfluidic chip that can adhere and conform to the leaf, collect guttation droplets whenever they come, and perform colorimetric detection of six kinds of chemical, 2) The Relay, a self-contained on-site camera module that can capture the colorimetric result from the chip, and 3) The Interpreter, an end-user app that can process the photos captured by the camera, identify the chemical levels, interpret the result, and provide suggested actions to the users.

\begin{figure}[b]
    \centering
    \includegraphics[width=1\linewidth]{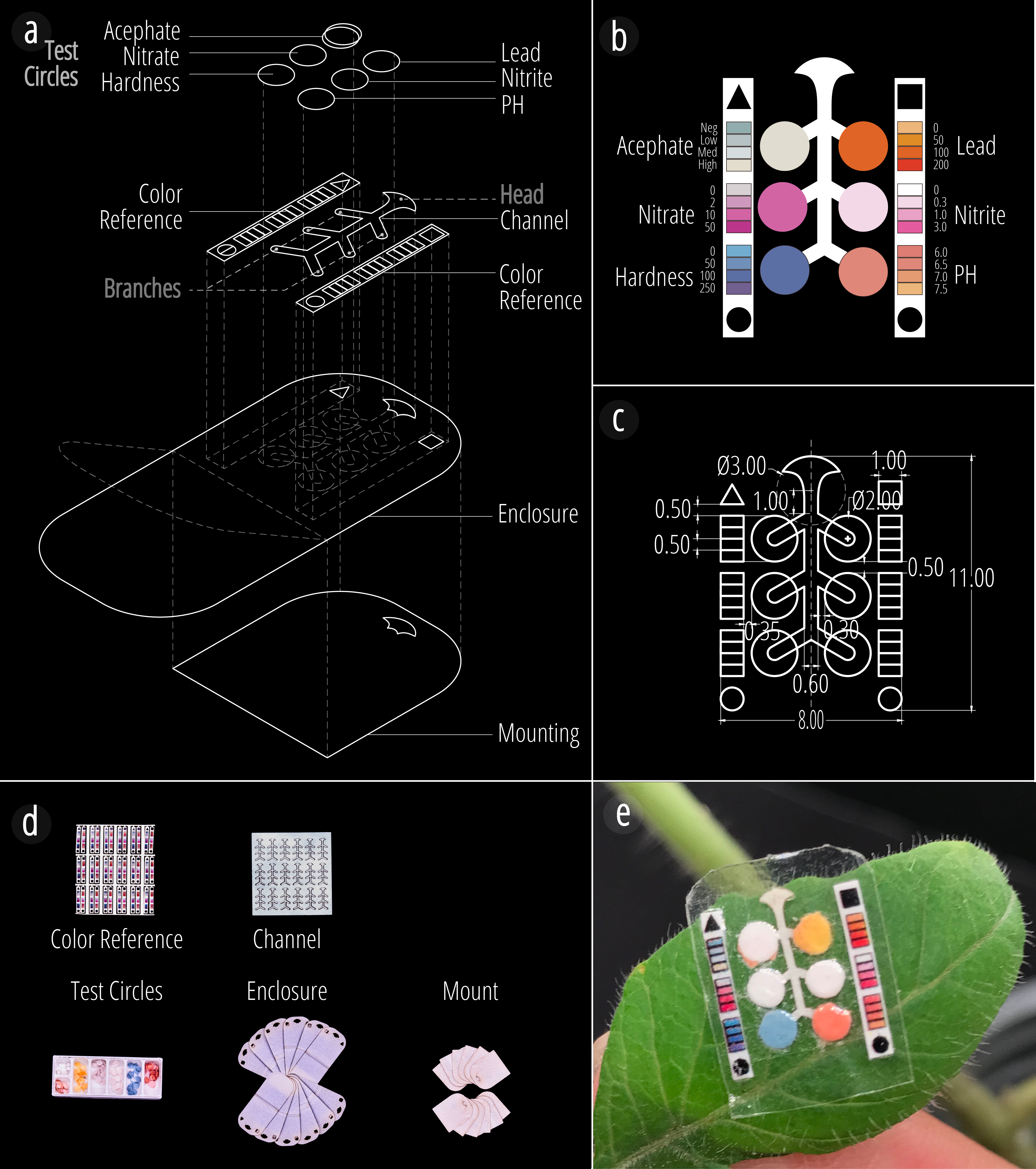}
    \caption{(a) Exploded axonometric diagram of the chip. (b) Color reference. (c) Dimensions of the chip (d) Components of the chip. (e) Chip implemented on the plant.}
    \label{fig:3}
\end{figure}

\subsection{The Sensor: Guttation Chip}
The colorimetric microfluidic guttation sensor is critical for guttation monitoring. It is compact, lightweight, low-cost, and can adhere and conform to the leaf in a manner that captures and routes guttation droplets through the microchannels to the test circles. We construct the guttation sensor based on a paper microfluidic chip rather than a 3D channel microfluidic chip (e.g., Polydimethylsiloxane cast chip, 3D printed chip) because

\begin{itemize}
    \item Paper microfluidic chips can work purely on capillarity, while 3D channel microfluidic chips require extra pressure at the inlet. For a sweat monitoring chip \cite{Koh2016, Zhang2019, Kwon2021}, an airtight inlet can be created when mounting the chip to the skin. Furthermore, perspiration can generate a natural pressure around 70 kPa \cite{Koh2016}. However, when we created an airtight inlet by enclosing the leaf edge and expected the guttation process to build positive pressure, we observed that the plant leaf guttating where it is enclosed. Therefore, we moved to a design that leverages paper microfluidic chips, which do not cause the same reaction.
    \item To avoid back-pressure that will impede the fluid flow, all channels must be interfaced to an outlet when designing a 3D channel microfluidic chip. Condensed moisture may get into and pollute the chip through such a structure. Paper microfluidic chips do not require such outlets. 
    \item Paper is very affordable and much easier to process. Moreover, assembling paper chips is very easy and rapid. No expensive and time-consuming instrument like a plasma treatment machine or SLA 3D printer is required.
\end{itemize}

\subsubsection{Designing the Guttation Chip}\

\textbf{Overview}. The guttation chips comprise a multi layer stack of five function components (Fig. \ref{fig:3}). From bottom to top, they are
\begin{enumerate}
    \item The mounting is a leaf-compatible double-sided adhesive layer with a sector shape opening for the guttation droplet collection;
    \item The enclosure is an ultra soft and thin one-sided waterproof adhesive with the same sector shape opening aligned with the opening in the mounting. In addition, it has a triangular cut and a rectangular cut to assist in locating the reference when assembling the chip;
    \item The reference consists of two printer paper strips with colorimetric reference to help identify the chemical concentration and image processing markers at both ends.
    \item The channel is a piece of tree shaped Japanese paper with one sector shape inlet and six branches. It can absorb the guttation droplet with the sector head and carry the fluid to the end of each branch;
    \item The test circles are for colorimetric detection of different chemicals, such as lead, nitrite, PH, acephate, etc.
\end{enumerate}

\textbf{Design Optimization.} We optimized the materials, geometry and function of the Guttation Chip’s design.

\underline{Materials}. For the Mounting, we use 3M 468MP Adhesive Transfer Tape, which has the just right amount of stickiness for mounting the chip and will not hurt the leaf when removing the chip. The enclosure needs to be lightweight, conformable and can protect the inside components from hard surfaces. We choose a thin, soft, waterproof polyurethane (PU) film (Areza Medical, Transparent Adhesive Film Dressing) for the enclosure. The reference is printed on regular printing paper. The test circles are punched from off-the-shelf test strips.

\begin{figure*}[t]
    \includegraphics[width=1\textwidth]{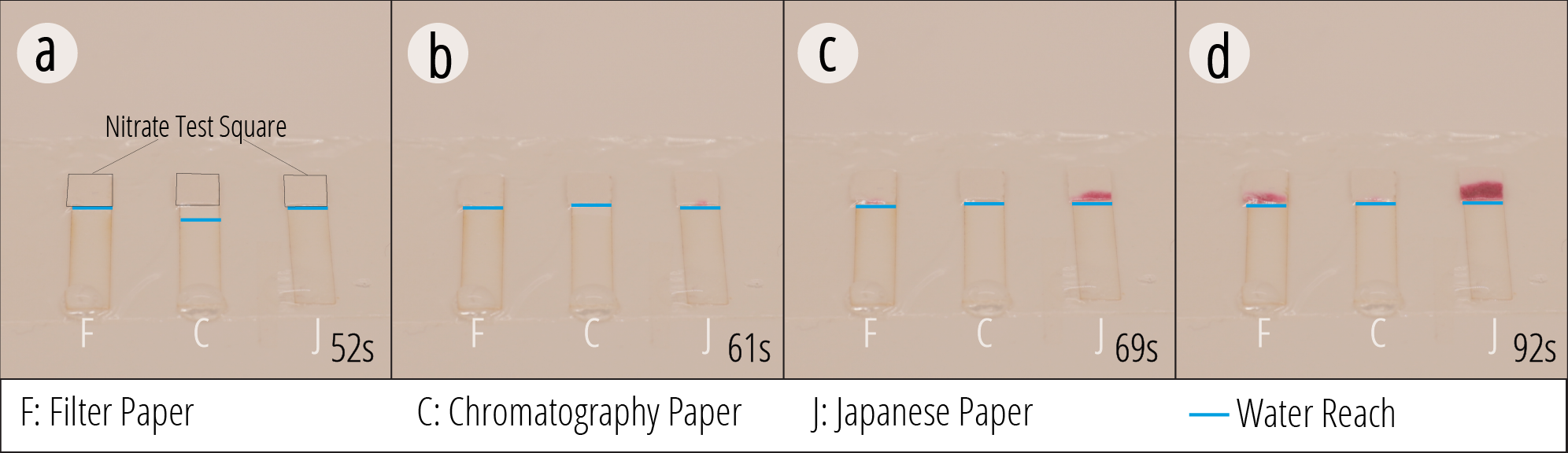}
    \caption{The experiment results for comparing the water absorbability and the chromatography property of different papers. (a) Water on filter paper and Japanese paper reaches the test squares (TS). (b) Water on chromatography paper reaches the TS; Nitrate reaches the TS on Japanese paper. (c) Nitrate reaches the TS on filter paper. (d) Nitrate reaches the TS on chromatography paper.}
    \label{fig:4}
\end{figure*}

As for the channel, we needed to find a paper material with good water absorption and minimal chromatography. Because chromatography will make solutes, especially large molecular weight solutes, fall behind the solution (in our case, water), accumulating and leading to inaccurate colorimetric test results. We select three kinds of paper with high absorbency: Japanese paper (ONAO), chromatography paper (LOSTRONAUT, Grade-1), and filter paper (Eisco Labs, medium speed - 85 GSM, 10 micron pore size). We then carry out an experiment to test them. All three papers are cut into 5 mm x 30 mm strips and have Nitrate test squares placed on one end. A sufficient and identical amount of Nitrate salt solution is dropped simultaneously to the other end of the strips. The time for water and Nitrate salt to reach the test square is recorded. As shown in Fig. \ref{fig:4}, All three papers have similar and very good absorbency, while the Japanese paper has the lowest chromatography property. Japanese paper is also much more affordable than the other two kinds of paper, making it an even better choice.

\underline{Geometry}. The geometry of the Guttation Chip is illustrated in Fig. \ref{fig:3}.c. Most of these parameters are established around the dimension of the test circle. The diameter of the test circle is 2 mm, which minimizes the sample volume requirement while not being too small to handle during manual chip assembly. The channel has close to the minimum width that most laser cutters can properly handle and has a relatively larger sector shape area to increase the chance of successfully collecting guttation droplets. A 0.3 mm, 0.3 mm, and 0.5 mm safe distance between the channel and the test circle, the reference and the test circle, and the test circles are reserved. The rectangle color reference patterns are grouped and aligned with the test circles. A guttation droplet’s volume can vary. For the tomato plants we have tested, the volume ranges from ~ 0.3 ml to ~ 5 ml. Even the smallest 0.3ml guttation droplet is far more than enough for our compact chip.

\begin{figure}[t]
    \centering
    \includegraphics[width=0.9\linewidth]{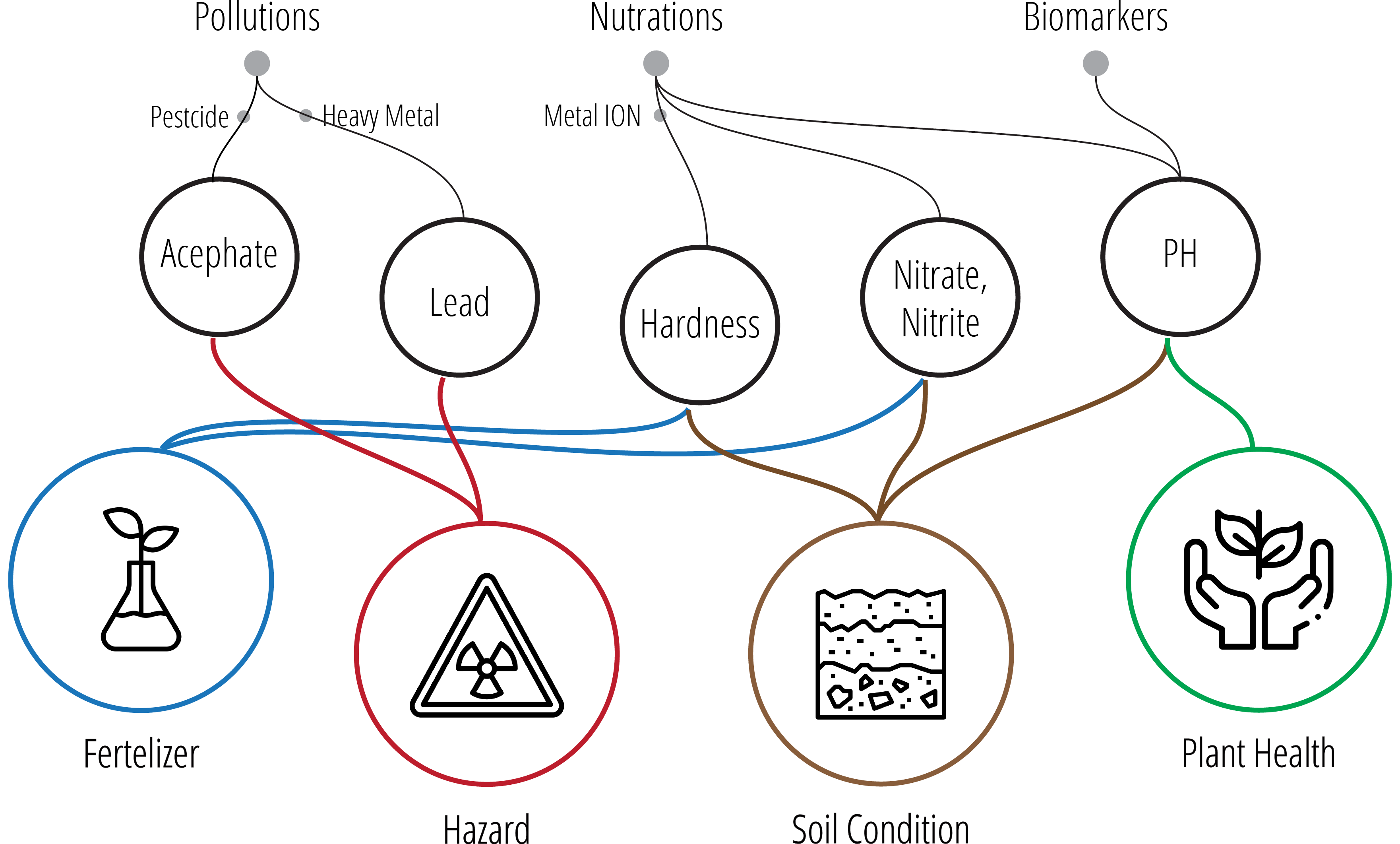}
    \caption{Chemicals the guttation chip can detect and what environmental factor they are related to}
    \label{fig:5}
\end{figure}

\underline{Function}. Some functional aspects have already been discussed, e.g. the channel has a larger inlet to facilitate guttation drop collection. In this part, we will mainly discuss how we design the core functional component of the chip: The colorimetric display. We select six chemicals to detect, with the following concerns and hypotheses (Fig. \ref{fig:5}):

\begin{itemize}
    \item Acephate: Pesticides pose a danger to the ecosystem by harming not only the insects they target, but also the plants and animals that consume them. Thus, we select one kind of pesticide, Acephate, to validate the chip’s capability of detecting pesticide residue. Acephate is an organophosphate systemic pesticide that is absorbed into plant tissues and sap where it is consumed by sap feeders. \cite{sharonm.douglasrichards.cowles}. Such systemic acephate can usually be detected in the guttation droplets, and can be hazardous to beneficial insects like bees who take guttation droplets as one kind of water/food source \cite{Kiljanek2016, Bonmatin2014}. 
    \item Lead: Land contamination is another problem that devastates our ecosystems. It is the most common type of contaminant in urban soil \cite{Brown2016} and is harmful to both animals and plants. Under the CDC guidelines, it is crucial to avoid agriculture on land with contaminated soil. Thus, we chose lead to see how soil lead pollution may be reflected in guttation droplets.
    \item Nitrate and Nitrite: Nitrogen fertilizer is one of the most widely used in modern agricultural practices, and it often comes as Nitrate salt \cite{Ware1988}. Monitoring Nitrate levels can help us understand if the plant is lacking Nitrogen. And if Nitrogen fertilizer is applied, this reading can also potentially help us understand if the appropriate amount of fertilizer is applied. While the natural level of Nitrite in soil is usually very low, Nitrate in plants can be converted into nitrite when concentrations are high, potentially indicating an overdose from nitrogen fertilizer. Nitrite is toxic to plants and harms their roots. Additionally, plants with high amounts of nitrite may not be safe for agriculture.
    \\Both Nitrate and Nitrite is harmful to humans, by disrupting formation of methemoglobin, limiting oxygen transport and causing anoxia. This dangerous hemoglobin causes oxygen starvation and is especially dangerous to young children \cite{Fewtrell2004}. They react with amines to form cancer-causing N-nitroso compounds \cite{Gangolli1994}. 
    \item pH: Plant sap pH level has been shown to be highly related to plant heath \cite{Verhage2021}. PH level fluctuations can be early warning signs of pests or disease susceptibility due to changes in soil condition. Since guttation droplets are comprised partially of xylem sap, we hypothesize that pH level fluctuations in guttation droplets can be observed when changes in soil pH occur.
    \item Hardness: The hardness level is an indicator of the calcium and magnesium ion concentrations in liquids. Calcium and magnesium are vital minerals necessary for humans and plants to remain healthy \cite{Sengupta2013}. Thus, it is important to monitor them.
\end{itemize}
 
We purchased many different kinds of test strips, prepared some standard solutions of the aforementioned chemicals, tested the chemicals with the strips, and selected the most accurate and sensitive one. 

The Reference color is extracted from the reference provided by the aforementioned product. As to the range, Lead, Nitrate, and Nitrite are determined by the EPA standard which is 200 ppm (lead in bare soil in play areas), 10 ppm, and 1 ppm respectively\cite{epanitrate}. The pH reference is selected based on the fact that most plant sap is weakly acidic when healthy. The Hardness reference is decided by making sure it can cover soft, medium and hard. The acephate test strip only comes with three levels of it detects - negative, low and high, to enable a more precise tracking of acephated, we add another medium level color fall between high and low.

\begin{figure}[t]
    \centering
    \includegraphics[width=1\linewidth]{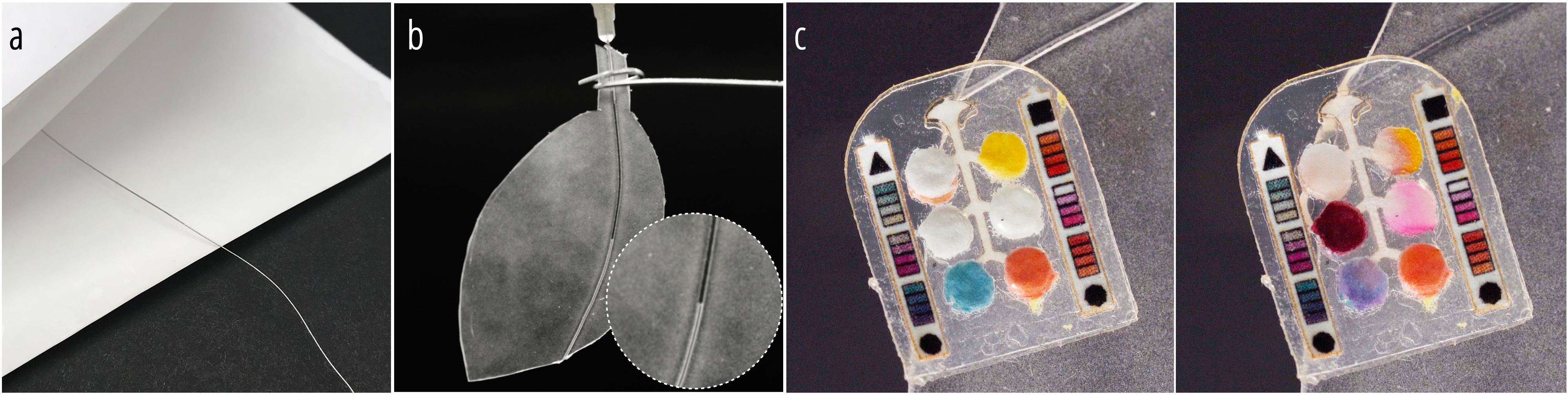}
    \caption{The bionic test bed. (a) The fabrication of the test bed. (b) Pumping a sample into the test bed (visualized with dyed water). (c, d) Pumping solution that contains acephate, lead, nitrate, nitrite, and calcium ions. The chip collects the sample and performs the color change.}
    \label{fig:6}
\end{figure}

\underline{Quick iteration}. To quickly verify and test our chip design, we implement a bionic test bed that mimics the plant guttation process (Fig. \ref{fig:6}). The artificial leaf is fabricated by heat pressing two layers of TPU film with a thin wire embedded. The wire then is removed, leaving a thin channel structure. Tubing with a needle head used to pump solution samples into the leaf is inserted into the thin channel. Fig. \ref{fig:6}.c. shows our chip can effectively collect guttation droplets and perform colorimetric detection.

\begin{figure}[t]
    \centering
    \includegraphics[width=1\linewidth]{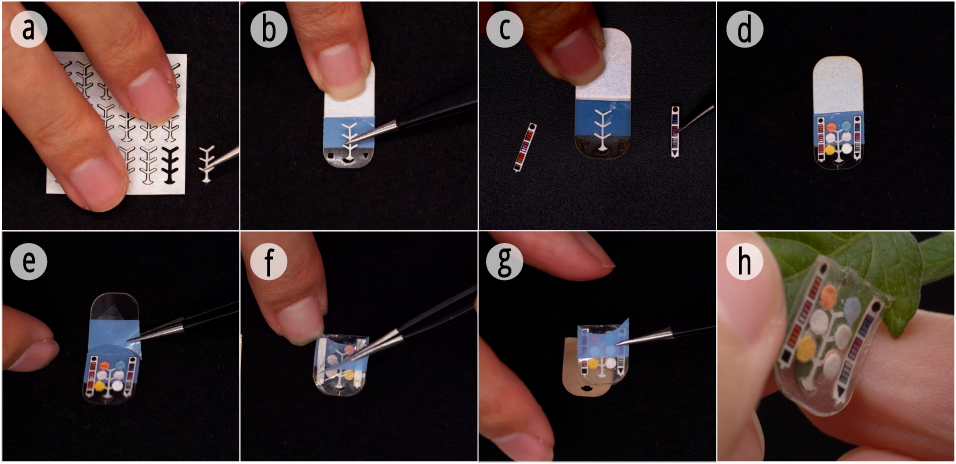}
    \caption{The assembling procedure of the guttation chip}
    \label{fig:7}
\end{figure}

\subsubsection{Making the Guttation Chip}
The channel (array), the reference (array), the mounting, and the enclosure are modeled with AutoCAD and Illustrator, then cut with a laser cutter. The test circles are punched out manually.

As shown in Fig. \ref{fig:7}, The chip is assembled with the following procedure: a). Get one channel from the array; b). Remove half of the protective cover of the enclosure and mount the channel. Make sure the sector shape parts are aligned; c). Mount the reference strips to the enclosure, leveraging the triangular and rectangular holes on the enclosure to align; d). Put the test circles in the right order, and note that the Acephate requires two overlapping test circles; e) Remove the other half of the protective cover and fold the enclosure; f). Remove one of the rigid support covers of the enclosure on the other side, and adhere to the mounting layer; g). Remove the other rigid support cover from the enclosure; h). Remove the rest of the protective cover from the mounting before adhering the chip to the plant.
\subsection{The Relay: On-site Camera Module}
\begin{figure*}[t]
    \includegraphics[width=\textwidth]{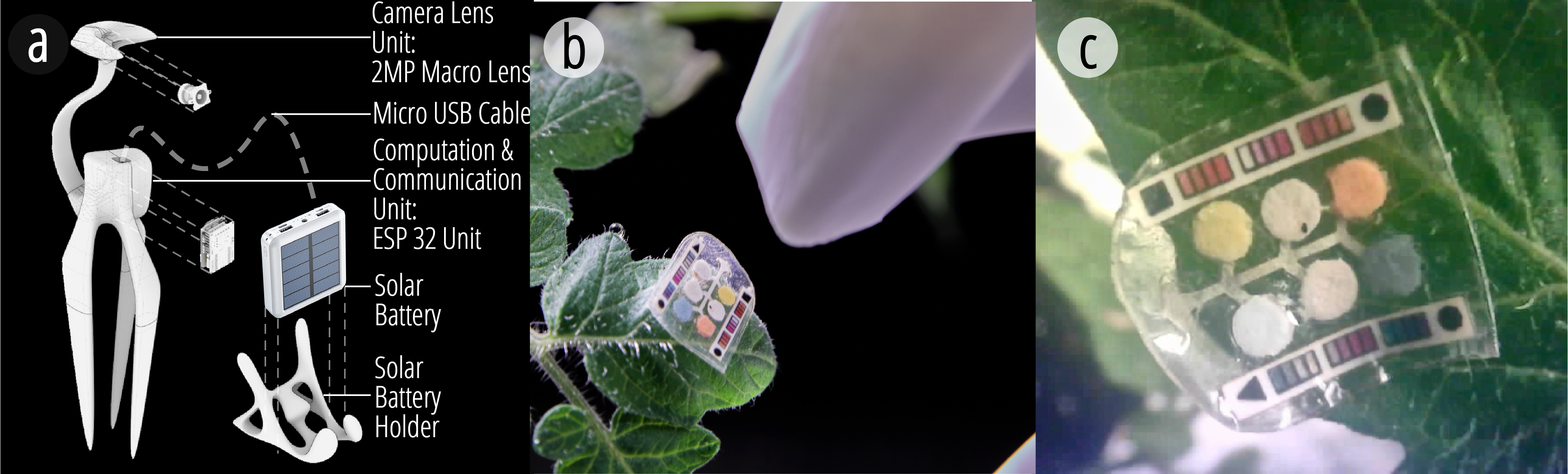}
    \caption{(a) The on-site camera module, (b)The chip captured by the camera. (c) Plant with a chip on it}
    \label{fig:8}
\end{figure*}

After assembly, the chip can be deployed on the leaf top. The inlet opening should face down and align with the end of the vein at the edge of the leaf. Younger leaves will guttate more than older leaves and are better candidates to deploy the chip. Guttation might not occur every day and when it occurs can be unpredictable. So, we develop an on-site solar-powered camera module to continuously capture the guttation chip (Fig. \ref{fig:8}). The camera module is programmed to take a picture of the chip every 15 minutes and wirelessly upload it to a laptop computer which is deployed as a local server.

The camera module has an ESP32 computation and communication unit (HK-ESP32-CAM-MB). It supports both Wi-Fi and Bluetooth dual-mode connectivity and features working power consumption as low as 30 mW. It is encased in a tripedal shell to be inserted into the soil. An OV2640 camera with an 80-degree lens is connected to the ESP32 board to take macro shots of the chip. A 12000 mAh solar power bank is used to power the system. The power bank is placed on a holder that tilts it at an angle of thirty degrees to face the sunlight. To feature a user-friendly and futuristic design, the case of the camera module is designed with reference to sci-fi animals. The lens case makes it look like it is staring at the chip on the leaf. Its streamlined body stands robustly and gracefully above the soil, like a fantastical beast attracted by the shape of the leaf veins.  

\subsection{The Interpreter: End-user Software}

\begin{figure}[t]
    \centering
    \includegraphics[width=1\linewidth]{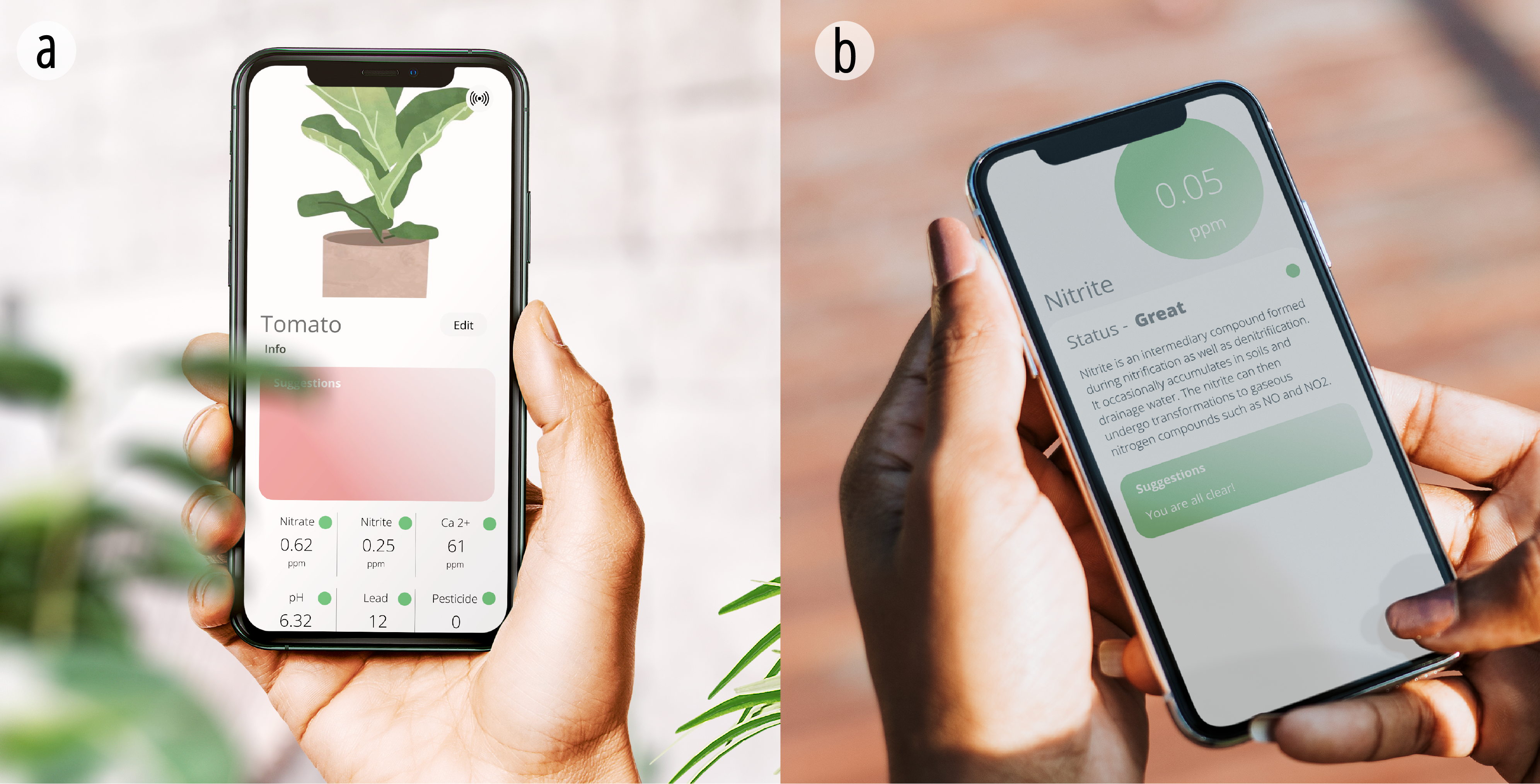}
    \caption{The end-user application main page. (a) The main page. (b) Chemical information page.}
    \label{fig:9}
\end{figure}

We developed an image processing algorithm and end-user app to get the concentration level of chemicals based on the chip photo and interpret the results.

After the photo is uploaded to the server, it is fed into the algorithm. The algorithm has three main parts:
\begin{enumerate}
    \item Identifying and isolating the chip from the entire image.
    \item Locating and extracting the colors of the six reactions and six corresponding tetrads of reference color bars.
    \item Using multivariate regression to get the concentration value based on the RGB values of each reaction and the four reference color bars on both sides of the chip. 
\end{enumerate}

We use OpenCV \cite{opencv_library} for color extraction and Scikit-learn \cite{scikit-learn} for regression. We first use the SIFT\cite{scikit-learn} algorithm to crop the chip from the photo and remap it to the X-Y domain by affine transformation. As the chip might be warped, the positions of the reference color bars and reaction circles might be shifted. Therefore, we further locate the bars and circles by detecting the location of the triangle, square and circle shapes on the four corners and finding their relative positions. This leaves us with six mutually exclusive sets, each comprising four reference colors and the colors from the center of the corresponding circle, which contains the chemical reaction.

For each of the six sets, we have the four ordered RGB values of the reference and one reactant RGB value.We fit a curve in a three-dimensional space according to the four RGB values with multivariate regression. We find the closest point on the curve to the reactant RGB value and interpolate the concentration value based on the four reference concentration values and the point location. The interpretation, which is preset based on our on-plant experiment presented in the next section, is selected according to the chemical detection result.

After the server finishes the image processing, the chemical concentration values and interpretations are sent to an end-user app. The app will show the user the data and interpretations, and the user can react accordingly. Fig. \ref{fig:9} demonstrates the application screens. In the main screen, the user can get a comprehensive understanding of the plant’s health by viewing the chemical concentration values and signals, which, in the example, are all green, indicating the plant is in good health. If there is anything worth noticing, the application will push suggestions to the main screen so that the user can take the necessary actions. The application also educates the user about what those chemicals are if the user clicks on the concentration value displayed on the screen.

\section{GUTTATION ANALYSIS USING THE DEVICE}
\subsection{Quantitative Guttation Detection}

\begin{figure*}
    \includegraphics[width=\textwidth]{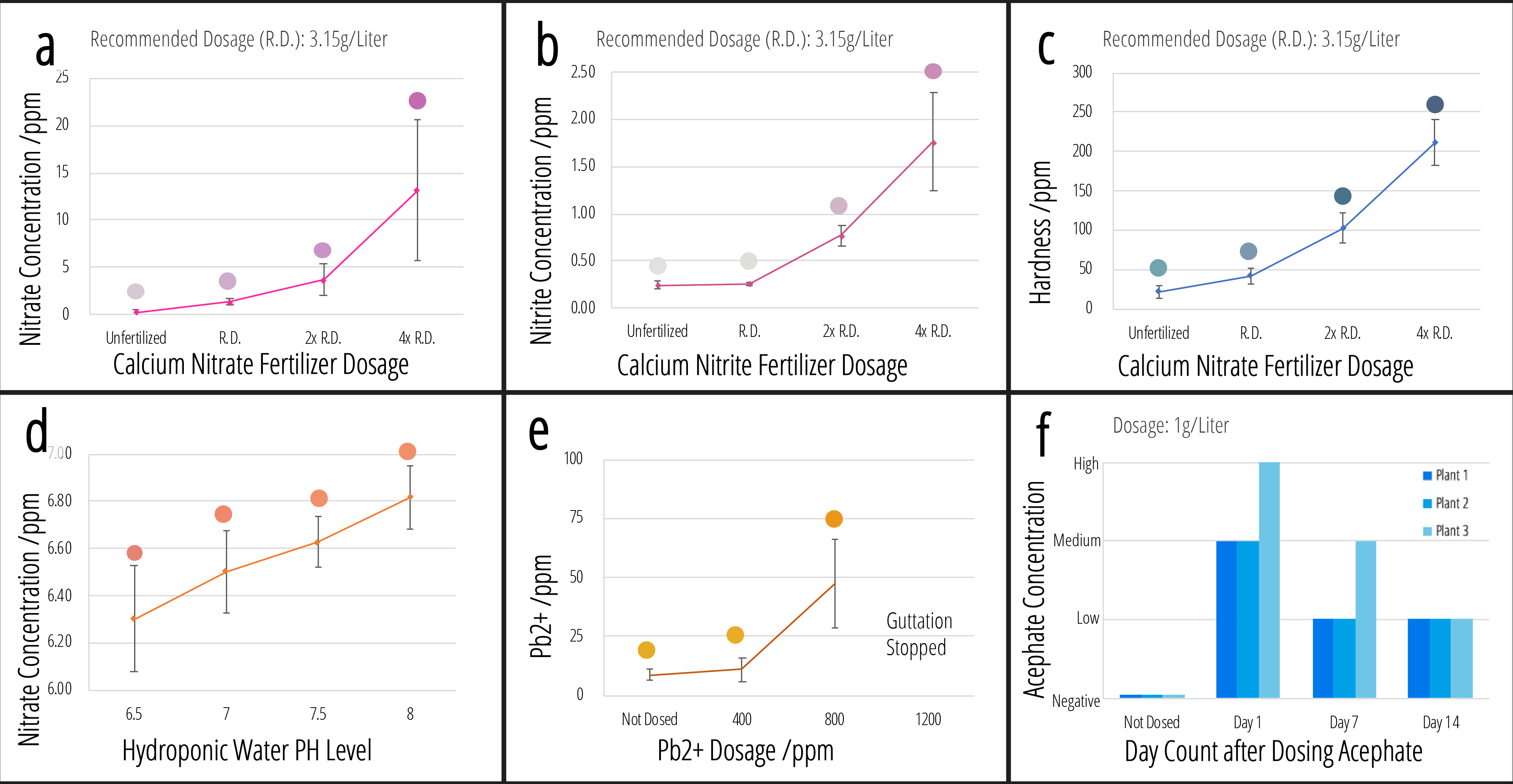}
    \caption{Quantitative guttation detection experiments using the chip. Nitrate(a), Nitrite(b), Hardness(c), pH(d), Lead(e), and Acephate(f)}
    \label{fig:10}
\end{figure*}

To validate our hypothesis on guttation fluctuation, we did a series of experiments with the Guttation Monitor system on one-month old tomato plants in a grow tent. The artificial sunlight was set to 8 hours at 25000 lux, and the temperature is set at 26$^\circ$C, and Humidity level is controlled at 60\%.
To precisely control the concentration of the chemicals of interest, we transferred the tomato plants to hydroponic systems with 1L filtered water, and let them sit for three days. We then deployed the Guttation Monitor system and applied chemicals of interest with various concentrations to the water. The next day (except the acephate which we monitored continuously for 14 days), the plants were kept at 90\% humidity, 26$^\circ$C for 3 hours in the early morning before the artificial sunlight cycle in order to induce guttation. 

For each chemical, data was collected from three chips deployed on three plants separately at different concentration levels:
Nitrate, Nitrite, and Hardness levels were controlled via Greenway Biotech calcium nitrate fertilizer. Fertilizer was used in the recommended dose of one tablespoon per plant (598.5 ppm Ca$^{2+}$, 456.75 ppm NO$_3^-$, 0 ppm NO$_2^-$), two times the recommended dose, and four times the recommended dose. A control group with no added fertilizer was also monitored.

Acephate by 97UP was used to monitor the effect of insecticides on guttation. The concentration of acephate in guttation was measured for two weeks after the initial dose of one gram per liter.

Lead nitrate salt was added to measure the lead concentration in guttation. Lead was added into the solution at 0 ppm, 400 ppm, 800 ppm, and 1200 ppm.

Standard 1 mol/L NaOH solutions are used to tune the hydroponic water pH level. The pH level started at 6.5 after initially leaving the plant in water for three days. Thus, we added NaOH solution to adjust the pH to 7, 7.5 and 8 to analyze the effect of the pH change on the plant roots and its impact on plant guttation.

The results are shown that the concentrations of the Nitrate and Nitrite increase as the dosages increase (Fig \ref{fig:10}.a, b). If the Nitrate is continually observed to be around 0, then the land may be barren and plants may need some Nitrogen fertilizer. When Nitrate concentration is detected to be around or above 5 ppm or Nitrite concentration is above 0.5 ppm, then we can tell the user probably overdosed the plant, and the user what they should consider. For example, removing excess fertilizer or providing extra water to dilute and wash away the excess fertilizer already seep into soil. If Nitrate is above 10 ppm \cite{Ward2018} or Nitrite is above 1 ppm, the user should be careful when eating (e.g. get some sample of the fruit and test before eating.), or wait until the Nitrate level drops back to normal range before harvesting the fruit.

The results also show that there is a positive correlation between hardness in the water and hardness in the guttation (Fig \ref{fig:10}.c). A hardness level of above 100 can be dangerous for plants. Thus our indicator allows users to identify hardness from excess calcium and magnesium in the soil.
There is also a positive correlation between pH in the soil and in guttation (Fig \ref{fig:10}.d). The optimal pH level of soil for tomatoes is between 6 and 7. If the pH level of guttation is detected to be too acidic or basic, then it could be an indicator that the soil pH is not optimal \cite{Astija2020}.

We found that plant guttation contains relatively low amounts of lead in comparison to the solution they are grown in (Fig \ref{fig:10}.e). This could be because tomato plants do not absorb much lead through the roots. Another cause could be because the plant does not expel the lead through guttation and thus, the lead ions are left in the plants’ tissue. When lead becomes detectable in guttation, above 25, then the concentration level of lead in the soil must be extremely high, around 800 ppm. The EPA guidelines recommend a maximum of 400 ppm lead in the soil for gardening use [cite]. Overdosing the plant with 1200 ppm lead caused the plant to cease guttating. Guttation only occurs in healthy leaves, thus a large increase in lead likely prevented the plant from undergoing guttation normally. Guttation is not very sensitive to lead, but our result tells us it can still be a convenient way to monitor extreme lead pollution. Detection of lead in guttation serves as a warning of lead pollution in the soil.

Acephate concentration was measured in levels of low, medium, and high. There was an obvious increase in acephate concentration after dosing the plants with one gram of acephate per Liter (Fig \ref{fig:10}.f). The acephate residue was detectable even two weeks after the initial exposure.

\subsection{Pilot Field Deployment Test}
\begin{figure}[b]
    \centering
    \includegraphics[width=1\linewidth]{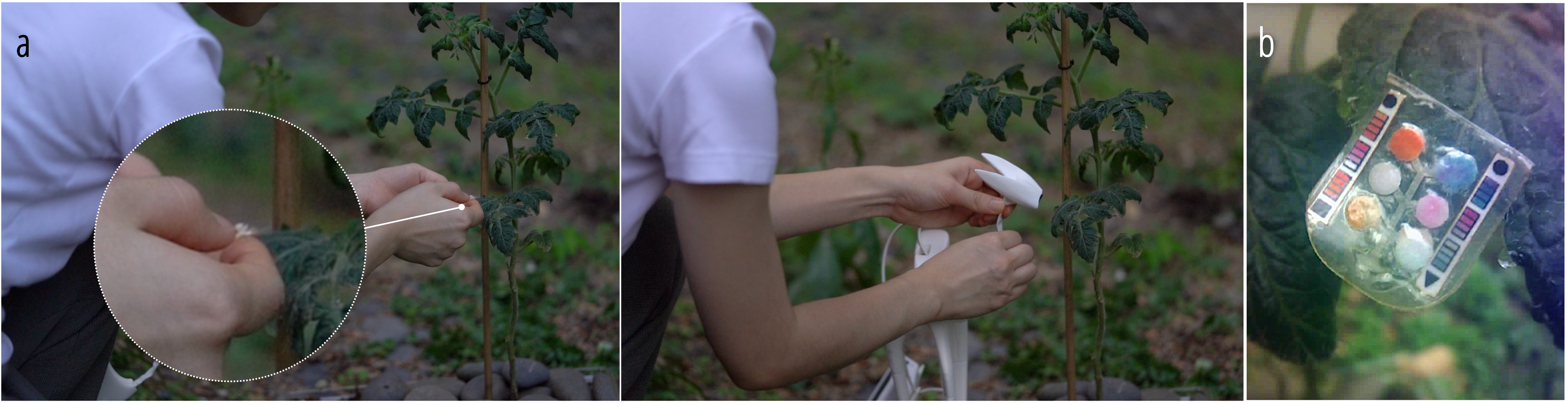}
    \caption{(a) The user is deploying the guttation chip and camera module at dusk, then waters the plant well. (b) Detection result captured by the camera on the next morning.}
    \label{fig:11}
\end{figure}
We deployed our system in a backyard garden to do a pilot validation of the system performance in the field (Fig. \ref{fig:11}). The guttation chip and camera module were deployed at dusk, after which the tomato plant was watered well. We then checked the results early the next morning. In about half of the cases, the chips were able to collect guttation and perform colorimetric reactions. The camera captured pictures of the chip after the sun rose. Adding a flash in the future could enable the camera module to capture photos at night as well.

\section{Identifying Design Opportunities Through Pilot Expert Interviews}

\subsection{Method}

\textcolor{black}{
To identify \textbf{new and unexplored applications} We identified a total of 15 subject matter experts (SMEs) in the United States with backgrounds spanning plant pathology, plant epidemiology, entomology, plant physiology, horticulture, (community) gardening, etc. Conversations with these experts around plant wearable technologies revealed that six of them (P1, P2, P4, P9, P10, P15) had more prior knowledge and experience with plant guttation. P1 and P9 were plant pathologists; P2, P4 and P10 had backgrounds in gardening, horticulture and plant physiology respectively; P15 was engaged in the operation of community gardens.} 

\textcolor{black}{We obtained signed nondisclosure agreements and informed consent forms from all 15 SMEs and conducted and recorded semi-structured interviews with them over Zoom. The goal of the interviews was to demonstrate our guttation monitor technology to them and seek their feedback on whether and how the technology could be valuable to them, and also to identify novel applications of our system. Our interviews also sought to answer broader questions on the applicability and desirability of plant sensing and actuation technologies. All participants were compensated with a 100-dollar gift card for their time and insights.} 

\subsection{Data Analysis}
\textcolor{black}{After cleaning up and anonymizing the interview transcripts, the authors qualitatively coded the data (excluding the initial presentation to the interviewees) using a thematic analysis approach. All 15 interview transcripts were quantitatively coded by 3 researchers using consensus coding and the codes were iteratively merged into themes around plant sensors and plant actuators.\textbf{While our interviews spanned a range of plant wearable technologies, here we will only present findings from six interviews (P1, P2, P4, P9, P10, P15) where data pertaining to guttation sensing was collected.}} 

\subsection{Results - Opportunities for Guttation Monitoring}

\begin{figure*}
    \includegraphics[width=\textwidth]{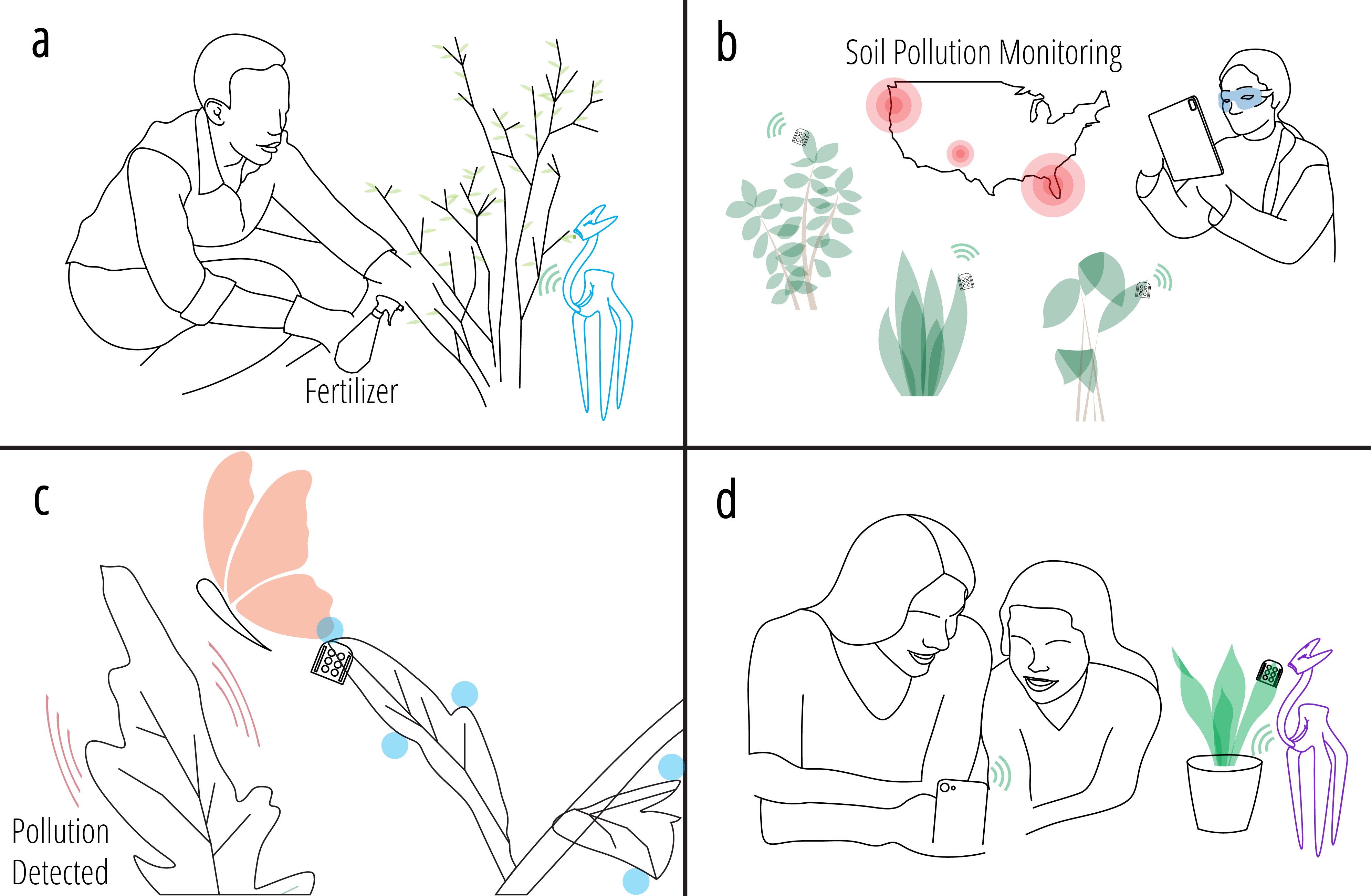}
    \caption{\textcolor{black}{(a) Improving one-to-one(s) human-plant(s) relationship, e.g., a novice gardener sprays fertilizer according to the suggestion from the device. (b) Improving human-plant relationships on a large scale, e.g., a scientist monitors the soil pollution and plant disease in the country via implemented chips in different states. (c) Bridging human-third species relationship, e.g., a shape-changing wearable warns the butterfly or kids when pollution is detected. (d) Bridging human-human relationship, e.g., a mother educates and discusses chemical and biological components detected with her daughter.}}
    \label{fig:12}
\end{figure*}

\textcolor{black}{Our interviews revealed that monitoring the chemical composition of plant guttation is meaningful in various ways. Firstly, it allows the users to perform plant health tracking, and can also help in preventing the onset and spread of diseases in plants. P1 highlighted that "you can pick up a lot of indicators of stress through the gut through what's excluded through the guttation, ...makes growing healthy plants much easier, especially for the unprofessional". When asked about the specific salts that might be important to detect through guttation, they explained "So I think it depends upon the plant, for example, you know, tomatoes are really susceptible to calcium deficiency. So you, you know, for those kinds of plants, you might want to monitor nitrogen plus calcium, or phosphorus or potassium, but for other plants, you know, that are more susceptible to boron such as our root crops, you know, you might, that might be a great opportunity to monitor boron content". We infer that expanding our current sensing capabilities to include other salts and choosing the salts based on plant species will further enhance the usefulness of our system.}  

\textcolor{black}{P1 also stressed upon the relationship between the chemical composition of guttation and disease resistance, positing that a healthy plant generally has a "good balance of chemicals" in the guttate, leading to a more diverse microbial community (near plant roots) and therefore reduced pathogen population.}

We learnt that guttation sensing could prove to be a valuable tool for plant epidemiology studies. Our experts \textcolor{black}{(P1, P9)} pointed out that the stomata, the epidermal pores that release guttation droplets, are also major entry points for plant pathogenic bacteria. Therefore, an analysis of the guttation fluid is useful in identifying plant pathogens well before they start showing signs of physical damage.
To quote P9:
\begin{quote}
    “To detect bacteria, for example, Ralstonia Solanacearum in tomatoes, we can cut the plant and put it in water and we see like the bacteria running out of the plant. So with guttation, I think we can do this too. The guttation will be already outside of the plant. And if we have sensors to detect some bacteria there it will be great because we'll be able to detect with the minimal amount of bacteria, and we will be able to detect sooner instead of later when you already have the symptoms. So I see an application for that.” 
\end{quote}

\textcolor{black}{P9 also stated that information collected through guttation analysis can allow users to not only prevent the spread of diseases (e.g. to your neighbor's garden, but also device mechanisms to build resistance without using pesticides or medicines: 
\begin{quote}
"As a scientist, I will try to relate information of these salts with the presence of the pathogen... So, basically, we can try to increase the levels of some of the salts to induce the resistance of the plant to fight against the pathogen without putting a lot of pesticides or other things there. I think the main thing that I'm thinking is basically to have sensors that do not damage the plant. So I think guttation sensing is a very interesting approach, where you can detect a lot of things. It's a natural process, and it's not invasive."
\end{quote}}
\textcolor{black}{While our guttation monitoring chip currently does not detect the presence of bacteria such as Ralstonia Solanacearum, this is possible theoretically and we think that this could be an exciting avenue for future guttation research using our non-invasive sensing platform.}

\textcolor{black}{To aid our understanding of how disease-related information and detection of other compounds like sugar in plants could be used to optimize grower output and commercial success, P2 explained:}
\begin{quote}
“So in terms of working with the growers, the plant health data that they want to know the most is sugar content, for a lot of things, a lot of the different crops that we work with. It's mainly sugar content, yield, and quality of the product at the end. And the best example I have of that is with snap beans, there is a 2\% threshold for white mold infection. And if it's above 2\%, the processing industry will turn back any shipment that is delivered to them. So it's very important to control those diseases early on, so that it doesn't get to the harvest stage where a grower will put in so many inputs throughout the season and then not be able to sell it.”
\end{quote}
\textcolor{black}{Similarly, P10 and P15 stressed how guttation could be helpful for studying and preventing over-fertilization in crops.} 

\textcolor{black}{P15 also provided valuable feedback from other aspects. P15 mentioned that people participating in (community) gardening could be exposed to soil contamination that can harm their health. However, "...with your sensors, we can do cheap soil tests instead of collecting and sending the samples to a lab". Besides, "...the roots of many plants can grow very deep, analyzing guttation can help us understand the condition of deep soil easier". P15 also pointed out that they were running educational programs in school gardens, and our platform "...can be an interesting teaching tool to help kids intuitively understand how their cultivation practices might affect plants, and why".}

Finally, beyond the regular crops like tomatos, we also heard from SMEs that there is ample willingness to use technology and plant sensors for expensive crops like cannabis that are currently grown indoors under artificial lights and using very intensive production systems. Describing how guttation sensing could be useful to cannabis production, P4 highlighted: 
\begin{quote}
“They're interested in THC production, the active ingredient. And so, some of these sensors for guttation could actually be used on the flower head of a cannabis plant. Because if you can find a close up image of the cannabis flower that has these little droplets of THC associated with the flower head. And so that's a, you know, that's a chemical that's a liquid, essentially, that is associated with the flower production. And so this technology could have an implication for instance, when you're using this to gauge the amount of THC that's being produced on a particular cannabis flower. That is a very interesting application.”
\end{quote}

\begin{figure*}
    \includegraphics[width=\textwidth]{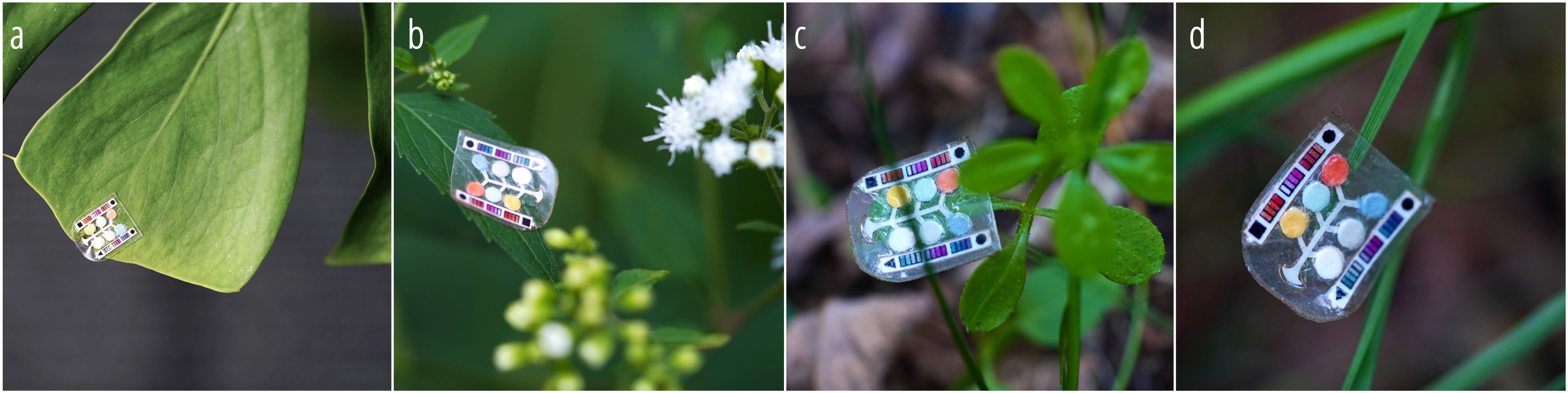}
    \caption{The guttation chip can be mounted to and be held on various plant leaves. From the large and thick \textit{Monstera deliciosa} leaf (a), to thin and sharp grass leaf (d).}
    \label{fig:13}
\end{figure*}

\subsection{Reflection - Potential Application Domains}

\textcolor{black}{After digesting the SMEs interview results, we proposed several potential application directions of our platform from the perspective of HCI researchers (Fig. \ref{fig:12}). }

\textcolor{black}{1) Improving one-to-one(s) human-plant(s) relationship. Our platform provides a novel solution for a better understanding of plant health and needs. We may leverage this to build education 
courses and tools \cite{rye2012elementary}. Moreover, though cultivation has been argued as a good way to relax, not everyone is good at taking care of plants. The stress associated with cultivating the plant may result in negative health, and well-being outcomes \cite{koay2020community}. We may build upon the technology to develop smart cultivation devices to facilitate cultivation (Fig. \ref{fig:12}.a). Lastly, we may develop novel and convenient research methods or tools for the plant scientists.}

\textcolor{black}{2) Improving human-plant relationships on a large scale. If widely deployed, with all the data it collects and the network it builds, our platform has the potential to be adopted to perform environmental quality monitoring, assist in urban and rural planning, etc. (Fig. \ref{fig:12}.b).}

\textcolor{black}{3) Bridging human-third species relationship. Human activities have greatly changed the environment. For example, small amounts of pesticide residues in plants that are harmless to humans may be fatal to pollinators \cite{blacquiere2012neonicotinoids}. Beyond the GUI interface for the human user we demonstrate, we can further integrate other output modalities like shape/color-changing that other species may appreciate (Fig. \ref{fig:12}.c).}

\textcolor{black}{4) Bridging human-human relationship. One may leverage our platform to prevent plant disease from spreading to neighbor's garden, or neighbors can compare their plants' micro-environment and learn from each other's planting experience; the plant status can also be shared with (remote) relatives, friends, or experts, promoting a better co-manage experience, etc.}

\section{DISCUSSION, LIMITATION, AND FUTURE WORK}

\subsection{Sustainability and Versatility}
The application target of plants poses unique criteria for the sensor chip, in terms of its sustainability and versatility in use. 

\underline{Sustainability}. The solar-powered camera unit can be repeatedly used, while the guttation chip, like many other microfluidic sensor chips, is single-use. But our paper-based chip is very low-cost and requires very little material to make, and can be easily disassembled and sorted before disposing. Seeking new material to replace the polymer Enclosure and Mounting can potentially make the entire chip biodegradable. 

\underline{Versatility}. 1) On the compatible plant types: The chip requires a very small sample to work and is very lightweight. Beyond tomato plants, which have an average-sized leaf, we have tried to mount the chip to various plants with good success (Fig. \ref{fig:13}). 2) On detecting other chemicals: \textcolor{black}{The chip itself is capable of being a carrier for any kinds of colorimetric test circles. Many off-the-shelf test circles for other chemicals, e.g., glucose\footnote{\href{https://www.homesciencetools.com/product/glucose-test-strips-50-pack/?srsltid=AYJSbAfl20hQCkkbTWXB8F0GO-Q90qfzJiK64y7plb84H95g354M50CvTsw}{Glucose Test Strips, Home Science Tools Inc}}, can be mounted to the chip.} However, if the test circle requires relatively strict conditions to react, the detection result may not be accurate. For example, when the temperature outdoors dropped to nearly 22$^\circ$C, we frequently got a false positive result for acephate in the field deployment test \textcolor{black}{while other test circles worked normally (Fig. \ref{fig:10}.c). It is because the test circles for acephate contain enzymes. Other limitations are the sensitivity of some test strips and the fact that for some chemicals there are currently no colorimetric test strips. For example, the lead test circles are not very sensitive to concentration level below 25 ppm, which result in the chip is not very responsive when the soil contains 400 ppm lead (the heavy metal level in guttation drops usually lower than in the soil).}

\textcolor{black}{Whereas, the good news is chemists have been devoted to studying indicators for making new test strips . For instance, the asparagine-modified AuNPs indicator for imidacloprid \cite{singh2022colorimetric}, a pesticide that is still widely used but highly toxic to bees; the indicator which can detect multiple toxic metal cations \cite{karthiga2013selective}; the indicator for salicylic acid \cite{tseng2014facile}, an important plant hormone; the ionogel based indicator for THC \cite{THC}; various indicators for detection of bacterial and fungal toxins \cite{toxin}; etc. Among them, we have tried to synthesize the imidacloprid indicator and the toxic metal cations indicator. Though we ultimately chose mature test strips that are approved for sale for more accurate results because the quality of the indicators we synthesized is not constant, we believe as additional types of test strips and indicators are introduced and mature, our chips can become more versatile by being able to detect more kinds of chemicals.}

\subsection{Further Iteration and Evaluation}
Our in vitro experiments demonstrate that our system can functionally work, and we believe our Guttation Monitor system can serve as a research and prototyping tool for HCI researchers and amateur, indoor/greenhouse gardeners. To push the technology further, we are planning to continue iterating on and evaluating our system by improving the following:

\underline{Iteration}. 1) Water Resistance: We find that our guttation chip can withstand light or normal dew. As dew tends to condensate on the leaf surface instead of the edge. However when there is heavy dew, dewdrops may gather, flow to the leaf edge, and seep into the chip through the opening. Also the chip channel will get wet if it is raining. One easy solution is to cover the whole leaf which has the chip mounted with a clear plastic bag. Or use a breathable and impervious adhesive membrane to create a waterproof chamber at the inlet opening. 2) Following our “End of the leaf veins and young leaves” principle will increase the chance of successfully collecting guttation droplets. However, due to the random nature of where the plant will guttate, we are still seeking an approach to increase the chance of collecting guttation. Adding a gathering channel along the leaf edge may increase the chance but the channel itself will waste some samples. Another potential approach is to see if we can leverage AI to predict where plants will guttate. 

\underline{Evaluation}. 1) To further evaluate the accuracy, we will conduct more on-plant experiments, and compare the concentrations of chemicals in guttation droplets obtained using our system versus laboratory-based analysis. 2) We have built connections with professional gardening organizations, and will conduct further in-field technical evaluations and user studies.

\section{Conclusion}
\textcolor{black}{This paper presented the Guttation Monitor, the first on-site and low-cost monitoring technology for guttation droplets. We introduced the development of the system. Then we carried out evaluations, including in vitro experiments, pilot in-field experiments, and experts interview. Based on the result, we speculated and discussed how this platform could be adopted by various users, including HCI researchers and designers, to develop technologies and design products that promote human-plant relationships in several directions.}

\bibliographystyle{ACM-Reference-Format}
\bibliography{guttation}
\end{document}